%% file: main.tex
\documentclass[journal]{IEEEtran}

\usepackage{xcolor,soul,framed} 

\colorlet{shadecolor}{yellow}
\usepackage[pdftex]{graphicx}
\graphicspath{{../pdf/}{../jpeg/}}
\DeclareGraphicsExtensions{.pdf,.jpeg,.png}
\usepackage{amssymb}
\usepackage[cmex10]{amsmath}
\usepackage{array}
\usepackage{mdwmath}
\usepackage{mdwtab}
\usepackage{eqparbox}
\usepackage{url}
\usepackage{algorithm}
\usepackage{algorithmic}
\usepackage{svg}
\usepackage{cite}
\usepackage{tikz}
\usepackage{textcomp}
\usepackage{hyperref}
\usepackage{lipsum}

\usepackage{eso-pic}
\begin{document}
\bstctlcite{IEEEexample:BSTcontrol}
    \title{Enhancement of High-definition Map Update Service Through Coverage-aware and Reinforcement Learning}
  \author{Jeffrey~Redondo,~\IEEEmembership{Student Member,~IEEE,}
      Zhenhui~Yuan,~\IEEEmembership{Member,~IEEE,}\\
      Nauman~Aslam,~\IEEEmembership{Member,~IEEE,}
}



\maketitle

\IEEEpubid{
\fbox{\parbox{\dimexpr\columnwidth-3\fboxsep-3\fboxrule\relax}{
    This work has been submitted to the IEEE for possible publication. Copyright may be transferred without notice, after which this version may no longer be accessible.
  }}
}
  
\input{abstract}

\IEEEpeerreviewmaketitle

\input{intro}

\input{related_work}

\input{annotation}

\input{problem_statement}

\input{design}

\input{table_simulation}

\input{results}

\input{conclusion}

\ifCLASSOPTIONcaptionsoff
  \newpage
\fi





\bibliographystyle{IEEEtran}
\bibliography{IEEEabrv,Bibliography}







\end{document}

%% file: abstract.tex
\begin{abstract}
    High-definition (HD) Map systems will play a pivotal role in advancing autonomous driving to a higher level, thanks to the significant improvement over traditional two-dimensional (2D) maps. Creating an HD Map requires a huge amount of on-road and off-road data. Typically, these raw datasets are collected and uploaded to cloud-based HD map service providers through vehicular networks. Nevertheless, there are challenges in transmitting the raw data over vehicular wireless channels due to the dynamic topology. As the number of vehicles increases, there is a detrimental impact on service quality, which acts as a barrier to a real-time HD Map system for collaborative driving in Autonomous Vehicles (AV). In this paper, to overcome network congestion, a Q-learning coverage-time-awareness algorithm is presented to optimize the quality of service for vehicular networks and HD map updates. The algorithm is evaluated in an environment that imitates a dynamic scenario where vehicles enter and leave. Results showed an improvement in latency for HD map data of $75\%$, $73\%$, and $10\%$ compared with IEEE802.11p without Quality of Service (QoS), IEEE802.11 with QoS, and IEEE802.11p with new access category (AC) for HD map, respectively.
    \end{abstract}

%% file: intro.tex
\section{Introduction} \label{intro}

\IEEEPARstart{I}{n} some developed countries, it is expected to see Autonomous Vehicles (AVs) hit the road by 2025 \cite{gov_uk, av_2025}. Currently, AVs already have a few self-driving features. However, to provide a completely autonomous driving experience in all scenarios as stipulated by the Society of Automotive Engineers (SAE) \cite{sae_level5} automation level-5 must be achieved. By reaching level 5, users with physical limitations would benefit from safe driving to hospital, work, and home \cite{disabilities_review,disabilities_expert,disabilities_policy}. In achieving the utmost level of automation, the utilization of HD maps has become crucial \cite{nvidea_hdmap} owing to its centimeter-level accuracy. Even more, it contains information that is not reachable by sensors \cite{hd_argument}.
\IEEEpubidadjcol
Regardless of the benefits of HD Map, and the actions proposed by researchers \cite{VRU} to the government to aid in the implementation of new legislation to support and accelerate the adaptation of AVs. A vehicular communication system plays a key role in generating HD maps. The key performance target is to support low-latency and heavy-data traffic applications to guarantee the safety of end users and vulnerable road users (VRUs).
\IEEEpubidadjcol
In high-density environments like Vehicular Ad-hoc Networks (VANET), wireless communication networks can become congested. To address this issue, there have been attempts to enhance the MAC protocol of the widely-used IEEE802.11p standard for VANET. These efforts aim to improve latency, throughput, packet delivery ratio, and handover \cite{sojourn_time_handover} through multiple access \cite{TDMA,HMAC}, new access category \cite{avaq_edca_new_ac,performance_analysis_HDMAP,low_latency_new_ac}, game theory solution \cite{adaptive_edca}, and Machine Learning (ML) algorithms such as Q-Learning \cite{q_learning_fairness}, etc. While these solutions have demonstrated improvements in Key Performance Indicators (KPIs) such as latency, throughput, and packet delivery, they have not yet taken into account the application of HD Map updates. Furthermore, their execution is not as simple as it might appear, as it entails making adjustments to existing standards, which may prove impractical. Therefore, there is still room for improvement in the current IEEE802.11p standard to reduce network deterioration, and maintain QoS.
\IEEEpubidadjcol
\subsection{Challenges and Objectives}
The de-facto Media Access Control (MAC) Layer in IEEE802.11p used in VANET suffers from packet collision in a high vehicular density due to the fixed Contention Window (CW) size as studied in \cite{ieee80211_cw}, and \cite{ieee80211_cw2}. In addition, the fairness problem is evident in cases where a few stations dominate the spectrum for a prolonged period of time, \cite{q_learning_fairness}. To mitigate these problems, new solutions have been proposed \cite{adaptive_edca}, \cite{q_learning_edca_policy_RL}; nevertheless, those solutions require modification to the standard which is undesirable. Hence, we aim to avoid any modification on the MAC layer. Subsequently, we propose a non-intrusive solution that works at the application layer and behaves as the backoff process and Enhanced Distributed Channel Access (EDCA) mechanism. Depending on the environment and coverage time, our solution assigns a waiting time before transmission to the wireless node. This approach addresses compatibility issues, making it applicable to any wireless communication system.

\IEEEpubidadjcol
Nevertheless, there is another challenge to overcome regarding the AVs' computational capacity. Due to the huge amount of data processing required for generating an HD Map that supports a fully autonomous driving experience; the control unit (CU) may become overloaded \cite{knowledge_driven} and its computing capacity may be insufficient \cite{chameleon}. Hence, the processing of HD Map is offloaded to cloud/edge/fog computing to accelerate the data processing. As illustrated in \cite{hdmap_processing_time},there is a 66\% enhancement observed  (reducing the processing time from 23 seconds to 7.8 seconds for a 20-meter HD Map.)

Most of those studies leverage the benefits of ML to improve the offloading process. However, this introduces an extra processing load on the On-board Unit (OBU) exceeding its computational capacity, given its current heavy real-time processing demands.

We have addressed the aforementioned challenges by implementing a centralized RL with a Q-Learning approach. Additionally, to reduce the number of states, we utilized the sojourn time, which calculation is based on four variables: velocity, speed, distance to the destination, and coverage area. 
Further details regarding the ML design and selection of states, actions, and rewards are discussed in the following sections.

\subsection{Contributions} 
\begin{itemize}
\item Ensuring QoS for HD Map sharing applications is a crucial undertaking. However, the proposed solutions require changes in the existing standard. To resolve this issue, this paper introduces a new QoS RL algorithm that is both non-intrusive and tunable. It is characterized by operating at the application layer, making it suitable for deployment across any wireless technology.
\item To reduce the number of states, and effectively extract the environment dynamics of a VANET network for the RL design. We developed for a basic discrete sojourn formula. This approach takes into account key variables like velocity, speed, distance, and coverage area, allowing the agent to effectively adapt to the ever-changing dynamics of the environment.
\item To provide the agent with a comprehensive understanding of the network and its performance. We decided on a reward function that encompasses the metrics latency, and throughput, plus the addition of bonuses, and penalties. By doing so, we enable the agent to make informed decisions. To mimic the EDCA mechanism used in IEEE802.11p, the inclusion of bonuses and penalties is particularly important as it assigns priorities to each service. This approach ensures that the network is optimized for all services, resulting in an improved QoS.
\end{itemize}

The paper is organized as follows: Section \ref{related_work} includes a comprehensive literature review on topics including the CW, EDCA, machine learning, utility function, and sojourn time. Section \ref{problem_statement} delves into the problem statement, while Section \ref{design} describes the design of the Q-learning algorithm. Section \ref{simulation} describes the simulation tools and assumptions used. Section \ref{results} presents the results and discussion. Finally, the paper concludes with section \ref{conclusion}.

%% file: related_work.tex
\section{Related Work} \label{related_work}
Several investigations have been conducted to improve IEEE 802.11 in the context of high-density and mobility wireless networks. These studies concentrate on adapting contention window (CW), developing queue management mechanisms, adapting EDCA, and integrating ML algorithms, all while optimizing a utility function. Additionally, there is an exploration of using sojourn time to enhance wireless communication. Nevertheless, most of them exclude the usage of HD map data. Besides, their implementation may call for further modifications to the standard.

\subsection{Contention Windows \& EDCA}
The investigations regarding CW and EDCA are divided into three subsections: high density, high mobility, and new access categories.

\subsubsection{High Density} One of the approaches to improve the Quality of Service (QoS) is to implement a queue management mechanism. For instance, mapping mechanisms between low and high-priority access categories (ACs) queues have shown a latency improvement of $13.8\%$ \cite{q_learning_edca_policy_RL}. Another option is a dynamic queue management that assigns data frames to an AC depending on the load \cite{dynamic_queue}. There are other similar works \cite{queue_IoV}, \cite{logical_EDCA}, and \cite{adaptive_edca}. Despite their improvement, the authors do not consider any other type of service than the standard EDCA (voice (VO), video (VI), best-effort (BE), and background (BK)). This could negatively impact new applications with low-latency requirements, as they will be assigned to the low-priority AC BE. When it comes to collaborative driving, having access to HD map data is crucial. As such, it's imperative to explore innovative strategies for accommodating new low-latency applications that might require this kind of data.

\subsubsection{High-Mobility} 
Congestion and throughput degradation are common in high-mobility VANETs. As a result, the network shows an increase in latency and a decrease in throughput and fairness. To improve the fairness of the network, authors in \cite{q_learning_fairness} developed a delay-awareness Q-learning solution that allocates CW, resulting in a $12\%$ improvement in fairness. Nonetheless, implementing it will necessitate adjustments to the existing standard. Additionally, there is not a direct understanding of mobility in the selection of the state space. It is solely based on the CW. As demonstrated in \cite{performance_analysis_HDMAP}, high mobility continues to be a problem that negatively influences network performance. Therefore, it is paramount to provide the RL algorithm with extra features that could provide a higher understanding of the vehicle's mobility.

\subsubsection{New Access Category}
An alternative approach to address low-latency applications involves creating a new access category. One proposal suggested a new AC for low-latency applications \cite{low_latency_new_ac}, while another suggested an adaptation of the current EDCA to a specific application such as video \cite{avaq_edca_new_ac}. Under this solution \cite{avaq_edca_new_ac}, each video resolution (low, HD, and 4K/8K) correspond to an AC. These solutions demonstrated improvements in low-latency applications. Even though, introducing a new AC for HD map data in VANET without considering a dynamic allocation for the CW is not adequate. Furthermore, such solutions would require modifications to the standard, making their implementation challenging for real-world applications.

The aforementioned studies highlighted the importance of having a dynamically tuned network, particularly given the inherent uncertainty caused by the high density and mobility of a VANET. Notably, HD Map a relatively new data type has not been involved as one of the main applications for prioritization within the framework of EDCA. Additionally, it is important to recognize that the proposed solutions require modifications to the existing standard, which adds a level of complexity to their implementation in real-world situations.

\subsection{Machine Learning and Utility Function}
Most recent studies leverage different RL algorithms such as Q-learning \cite{q_learning_fairness}, policy gradient \cite{q_learning_edca_policy_RL}, deep RL \cite{adaptive_cw}, \cite{cw_deep_RL_ieee802.11ax}, and long short-term memory (LSTM) prediction model \cite{sojourn_time_handover}, and game theory \cite{adaptive_edca} to reduce the latency. Depending on the application, other authors have proposed a utility function to maximize, for instance, resolution and latency \cite{chameleon}, or only latency by considering collision time \cite{adaptive_cw}. Nonetheless, the implementation of these solutions demands either a high computing capacity or a modification of the existing standard.

\subsection{Sojourn Time}
Researchers have shown the effectiveness of utilizing sojourn time (defined as the time an object will spend in a system) to enhance network performance. This is particularly important for reducing the number of handovers triggered by varying signal strength \cite{sojourn_time_handover}, \cite{sojourn_time_ultradense}. When users experience a ping-pong effect due to rapid signal fluctuations, unnecessary handovers and packet loss can occur,  ultimately resulting in increased latency. To address this issue, authors in \cite{sojourn_time_ultradense} has proposed an algorithm that decides which small cells are selected as target depending on the sojourn time threshold. The solution showed an improvement of $30\%$ in terms of handover failures. Our strategy adopts sojourn time in a distinct approach. Instead of selecting the target, the AV is instructed when to initiate transmission through the reception of the action waiting time. As a consequence, reducing the likelihood of unnecessary transmissions.

%% file: annotation.tex
\begin{table}
    \caption{Annotation Table}
    \centering
    \begin{tabular}{c|c|c|c}
    \hline
    Variable & Definition & Variable & Definition \\
    \hline
    $\alpha_1$, $\alpha_2$ & Coefficients & $Sj$ & Discrete Sojourn Time \\
    $\mathcal{A}$ & Action Set  & $sj$ & Sojourn Time \\
    $\mathcal{B}$& Bonus  & $slope$ & Slope for Direction \\
    $\mathcal{B}_\mathcal{L}$& Latency Bonus& $\mathcal{T}$ & Time Set \\
    $\mathcal{B}_\mathcal{R}$& Data rate Bonus & $Tv$& Veh Active at the RSU\\
    
    $\mathcal{C}$ & Category Set &  $Tcv$&  Veh Active per Category\\
    $d$ & Distance & $U$ & Utility Function \\
    $\mathcal{L}$& Latency & $w$& Waiting Time\\
    $\mathcal{L}_{max}$& Max Latency & $w_{max}$& Max Waiting Time\\
    $J$& Fairness Index & $x$& Binary Index\\
    $\mathcal{P}$& Penalty & $\mathcal{R}$ & Data Rate\\
    $\mathcal{P}_\mathcal{L}$ & Penalty for latency & $\mathcal{R}_{min}$& Min Data Rate \\
    $\mathcal{P}_\mathcal{R}$ & Penalty data rate & $\mathcal{V}$ & Vehicle Set\\
    
    $\pi$ & Policy & $\epsilon$ &e-greedy \\
    $S$& State Space &  &  \\
    \end{tabular}
    \label{tab:notation}
    \end{table}

%% file: problem_statement.tex
\section{Problem Statement} \label{problem_statement}
For this case, we assume a dynamic vehicular traffic flow with time step $t = \{1,..,T\}$ where a set of AVs $\mathcal{V} = \{1,...,N\}$ enters and leaves the environment, mimicking the dynamic traffic flow of an urban area. Each vehicle is assigned categories for data transmission, denoted by a set $\mathcal{C} = \{1,...,M\}$.

The goal is to maintain throughput and latency requirements for different types of services. Therefore, an utility function is considered \cite{adaptive_edca} that contains throughput ($\mathcal{R}$) and latency ($\mathcal{L}$). 

In our case, we employ penalties and bonuses according to the category to improve the control of the agent's learning, as represented by the following equation.

\begin{equation}
    \begin{split}
    U(c) = \alpha_1\frac{\mathcal{R}(c)}{\mathcal{R}_{max}(c)}- \alpha_2\frac{\mathcal{L}(c)}{\mathcal{L}_{max}(c)} + \mathcal{P} + \mathcal{B}
    \label{eq:utility_function_penalty}
    \end{split}
\end{equation}

The values of throughput and delay are normalized depending on the category $c \in \mathcal{C}$ (type of service) that is evaluated. In this way, utility values will be in the same range for each of $\mathcal{C}$, as their throughput and delay requirements differ. The coefficients $\alpha_1$, and $\alpha_2$ are weights to provide a trade-off between $\mathcal{R}$ and $\mathcal{L}$. 

Henceforth, the maximization problem for the RSU is formulated as follows:

\begin{equation}
    \underset{w_{v,t}}{\max}  \sum_{v \in \mathcal{V}} \sum_{t \in \mathcal{T}} x_{v,t} U_{v,t}(c) \quad ,\forall c \in \mathcal{C}, v \in \mathcal{V}, t \in \mathcal{T}
    \label{eq:max_problem}
\end{equation}

subject to:

\begin{equation}
    x_{v,t} \in \{0,1\}
    \label{eq:constraint_x}
\end{equation}
\begin{equation}
    \frac{1}{|\mathcal{V}|} \sum_{v=1}^{|\mathcal{V}|} \mathcal{L}_{v}(c) \leq \mathcal{L}_{\text{max}}(c) \quad , \mathcal{L} \in \mathbb{R}
    \label{eq:constraint_L}
\end{equation}
\begin{equation}
    \sum_{v=1}^{|\mathcal{V}|} \mathcal{R}_{v}(c) \geq \mathcal{R}_{\text{min}}(c) \quad , \mathcal{R} \in \mathbb{R}
    \label{eq:constraint_R}
\end{equation} 
\begin{equation}
    w_{v,t}(c) \leq w_{\text{max}}(c) \quad , w \in \mathbb{R} \quad , w \neq 0
    \label{eq:constraint_w}
\end{equation}

where $x_{v}$ is the binary index that could be either 0 or 1 indicating that a vehicle is allowed to transmit. Constraint (\ref{eq:constraint_L}) refers to the maximum tolerable latency per service type, (\ref{eq:constraint_R}) is the minimum data rate per service type, and (\ref{eq:constraint_w}) is the maximum waiting time allowed per category. 

This formulated maximization problem contains multiple variables and relationships increasing the complexity to solve it. The relation of $\mathcal{L}$ and $\mathcal{R}$ against CW and number of vehicles are described here:

\begin{equation*}
    \mathcal{L} \propto CW \quad and \quad \mathcal{L} \propto |\mathcal{V}|
\end{equation*}

\begin{equation*}
\mathcal{R} \propto CW^{-1} \quad and \quad \mathcal{R} \propto |\mathcal{V}|^{-1}
\end{equation*}

The CW, number of connected vehicles, and latency share a direct relationship, while throughput is inversely proportional. These exhibit non-linear dynamics. Additionally, the activation of the variable $x$ depends on $w$. Therefore, to maximize the utility function $U$, the optimal values of $x$ and $w$ must be found.

Solving this problem analytically is difficult due to the complex relationship and the potential non-linearity in the dynamic selection and increase of the CW value, coupled with changes in vehicle numbers resulting from traffic flow. To overcome this, an optimization algorithm capable of capturing complex patterns is necessary. In this regard, an RL algorithm is a suitable choice as it can adapt to non-linear relationships, unlike traditional analytical methods. The chosen design is a Q-learning algorithm whose task is to control the index $x_{v}$ by assigning a waiting time $w$ (\ref{eq:constraint_w}) to the vehicle $v$ for the next transmission. For ease of reference, Table \ref{tab:notation} provide a summary of the notations used in this paper.

%% file: design.tex
\section{Design} \label{design}
To develop a non-intrusive QoS solution to the MAC layer. We opted to design a solution that runs on the application layer. To this end, we have focused on designing a Q-Learning algorithm that utilizes coverage time (sojourn time) to reduce complexity. Additionally, to mimic and improve the QoS enhancement in IEEE 802.11p \cite{iee802_11p_standard}, actions and penalties are established accordingly.

\subsection{Reinforcement Learning}
As discussed previously, our solution utilizes less computational capability. To achieve this goal, the utilization of Q-learning temporal difference (TD), a model-free method, has been chosen to optimize the reinforcement learning (RL) algorithm. This optimization aims to enhance the delivery of HD Map updates in a vehicular network.

This approach combines the strengths of dynamic programming (DP) and Monte Carlo (MC) methods. Given the stochastic nature of the vehicular network, the transition probability $P(s,s')$ is unknown, making it imperative to determine the optimal policy $\pi$. To obtain $\pi$, we opted for off-policy TD control Q-learning, which directs the agent to seek out the optimal policy $\pi$.

At each step $t=\{0,....T\}$, the agent gets an observation of the environment and performs an action $a$ given by the $\pi(a,s)$ using $\epsilon$-$greedy$ epsilon. After the agent acts, a negative or positive reward $r$ is obtained, and the Q value table is updated using eq. (\ref{eq_q_learning}), that contains the value of the next state with a discount factor ($\gamma$), and a learning rate ($\alpha$). 

\begin{equation}
    Q(s,a) = Q(s,a) + \alpha\left[r+\gamma*max_{a'}Q(s',a')-Q(s,a)\right]
\label{eq_q_learning}
\end{equation}

To provide the agent with network performance awareness, the reward function is the utility function eq. (\ref{eq:utility_function_penalty}). For simplicity we denote $y = \gamma*max_{a'}Q(s',a')-Q(s,a)$. And, by substituting (\ref{eq:utility_function_penalty}) in (\ref{eq_q_learning}), the proceeding eq. (\ref{eq_q_learning}) becomes, 

\begin{equation}
\begin{split}
        Q(s,a) = Q(s,a) + \alpha[U + y]
    \label{eq_q_learning_2}
\end{split}
\end{equation}

\subsection{State, Action, Reward}
Moving on with the Q-learning design, the state, action, and reward are described in this subsection.

\subsubsection{State}
The state set $\mathcal{S}$ is defined as follows:

\begin{equation}
    S = \{Sj, Tv, C, Tcv\}
\end{equation}

$Sj$ being the sojourn time with values from zero to four as described further in this section. $Tv$ is the total number of vehicles active in the RSU with a maximum of $N$. $C$ the category with maximum $M$, and $Tcv$ the total number of actives vehicles per category $c$.

\begin{equation*}
    Sj = \{0, 1, 2, 3, 4\}
\end{equation*}
\begin{equation*}
    Tv = \{0,...,N\}
\end{equation*}
\begin{equation*}
    \mathcal{C} = \{0,...,M\}
\end{equation*}
\begin{equation*}
    Tcv = \{0,...,N\}
\end{equation*}

\paragraph{Categories}
Each vehicle is assigned one category to transmit during the simulation, denoted by $\mathcal{C}=\{VO, VI, BE, HD\}$.

\paragraph{Sojourn time}

Incorporating sojourn time into our RL algorithm yields the following advantages. Firstly, as discussed in \cite{sojourn_time_handover}, \cite{performance_analysis_HDMAP}, \cite{q_learning_fairness}, the topology of the network is affected by the uncertainty of the traffic flow (density and mobility). Therefore, by using the sojourn time, we can perceive the behavior of the traffic flow and react before the network performance deteriorates. Secondly, as stipulated in\cite{sojourn_time_ultradense}, sojourn time is closely linked to velocity. Consequently, we reduce the state space by combining speed, location, and direction.

To calculate sojourn time, we have included the coverage area of the RSU, which we have set to $200m$. The rationale was based on previous research \cite{coverage_range}, and \cite{coverage_range_1} that has shown a decline in throughput, packet reception, and latency beyond the $200m-300m$ range for IEEE802.11p. Nevertheless, this value could be adjusted by the stakeholders depending on the wireless technology.

The sojourn time is calculated as follows. The RSU edge server gets the velocity and location of the vehicle. 

Thus, the distance between RSU and AV is calculated with the Euclidean distance formula. Then the direction of the vehicle is calculated with the $slope = \Delta y/\Delta x$. Finally, the sojourn time formula is,

\begin{equation}
{sj} = 
    \begin{cases}
     \frac{(area/2) - d}{speed} ,&{\text{if moving towards the RSU}} \\ 
    {0}, &{\text{if moving away from the RSU}} 
\end{cases}
\end{equation}

As we have adopted for a Q-learning approach, it is essential for the state space to be discrete. In order to determine the range of each discrete sojourn time, we began by calculating the maximum amount of time a vehicle could spend within the coverage area, taking into account the maximum speed limit in an urban area of $17$m/s. We then divided this value by the total number of discrete values, five, to arrive at our desired range, as outlined in (\ref{sojourn_range}).

\begin{equation}\label{sojourn_range}
S_j =
\begin{cases}
    0, & 0 \leq sj \leq 2.9 \\
    1, & 2.9 < sj \leq 5.6 \\
    2, & 5.6 < sj \leq 8.4 \\
    3, & 8.4 < sj \leq 11.2 \\
    4, & sj > 11.2 \\
\end{cases}
\end{equation}

\paragraph{Active User Counts}
To record the number of vehicles active in the RSU and per category. We have developed two short subroutines (\ref{algo:ActiveUserCaount}), and (\ref{algo:checkActiveUsers}). The first subroutine registers the user upon receiving a packet, while the second subroutine, scheduled to run every second, validates user activity. Inactivity is determined if the RSU fails to receive a packet within a ten-second period.

\begin{algorithm}
\caption{Active Users Count Subroutine}
\label{algo:ActiveUserCaount}
\begin{algorithmic}[1]
\STATE Initialize map TotalActiveUsers
\
\WHILE{packet received}
  \STATE Register the user Id and category
  \STATE start counter with 10s
  \IF{checkActiveUser not active}
    \STATE schedule subroutine checkActiveUsers every 1s 
  \ELSE
    \STATE \textbf{break}
  \ENDIF
\ENDWHILE
\STATE \textbf{Finish the subroutine}
\end{algorithmic}
\end{algorithm}

\begin{algorithm}
\caption{checkActiveUsers Subroutine}
\label{algo:checkActiveUsers}
\begin{algorithmic}[1]
\STATE find user
\STATE get counter
\IF{User found} 
    \STATE decrease counter by 1
    \IF{counter == 0}
    \STATE delete user from map TotalActiveUsers
    \ENDIF
\ENDIF
\end{algorithmic}
\end{algorithm}


\subsubsection{Actions}
In order to ensure Quality of Service (QoS) for the system, a set of actions $\mathcal{A} = \{0,..,k\}$ has been chosen, where $k$ represents the final action that can be taken within the set. The mapping of these actions for each $c$ is represented by eq. (\ref{eq:map_actions}).

\begin{equation}
    w(c) = a \cdot \left(\frac{w_{max}(c)}{|\mathcal{A}|}\right)
    \label{eq:map_actions}
\end{equation}

\subsubsection{Reward}
In accordance with the aforementioned, the utility function (\ref{eq:utility_function_penalty}) will be utilized to determine the reward, while also outlining the constraints for each category and the corresponding penalties and bonuses.

In order to achieve optimal results, it is important to carefully consider the unique requirements of each category. Therefore, by selecting the appropriate QoS configuration and adjusting penalties and bonuses, we can prioritize certain services and ensure that the most important categories receive the highest level of attention, see eq. (\ref{eq:penalties_rate_VO_VI_HD}), (\ref{eq:penalties_BE1}), (\ref{eq:bonus_Latency_VO_VI_HD_BE}), and (\ref{eq:bonus_Latency_BE}).

Penalties are defined as follow:
\begin{subequations}
\begin{align}
\label{eq:penalties_rate_VO_VI_HD}
\mathcal{P}_{VO,VI,HD} &=
\begin{cases}
    \text{if }\mathcal{L} > \mathcal{L}_{max} \text{ then }\mathcal{P}_\mathcal{L} \\
    \text{if } \mathcal{R} < \mathcal{R}_{min}\text{ then }\mathcal{P}_\mathcal{R}  \\
\end{cases} \\
\label{eq:penalties_BE1}
\mathcal{P}_{BE} &=
\begin{cases}
   \text{if } \ \mathcal{L} < \mathcal{L}_{max} \text{ then }\mathcal{P}_\mathcal{L} \\
     \text{if } \ \mathcal{R} > \mathcal{R}_{min} \text{ then } \mathcal{P}_\mathcal{R}
\end{cases}
\end{align}
\end{subequations}

And the Bonuses as:
\begin{subequations}
\begin{align}
    \label{eq:bonus_Latency_VO_VI_HD_BE}
    \mathcal{B}_{VO,VI,HD} &=
    \begin{cases}
       \text{if }\mathcal{L} \leq \mathcal{L}_{max} \text{ then } \mathcal{B}_\mathcal{L} \\
       \text{if }\mathcal{R} \geq \mathcal{R}_{min} \text{ then } \mathcal{B}_\mathcal{R}
    \end{cases} \\
    \label{eq:bonus_Latency_BE}
    \mathcal{B}_{BE} &=
    \begin{cases}
       \text{if }\mathcal{L} \geq \mathcal{L}_{max} \text{ then } \mathcal{B}_\mathcal{L} \\
       \text{if }\mathcal{R} \leq \mathcal{R}_{min} \text{ then } \mathcal{B}_\mathcal{R}
    \end{cases}
\end{align}
\end{subequations}

\subsection{Algorithm}
\begin{figure}[h]
  \begin{center}
  \includegraphics[width=3in]{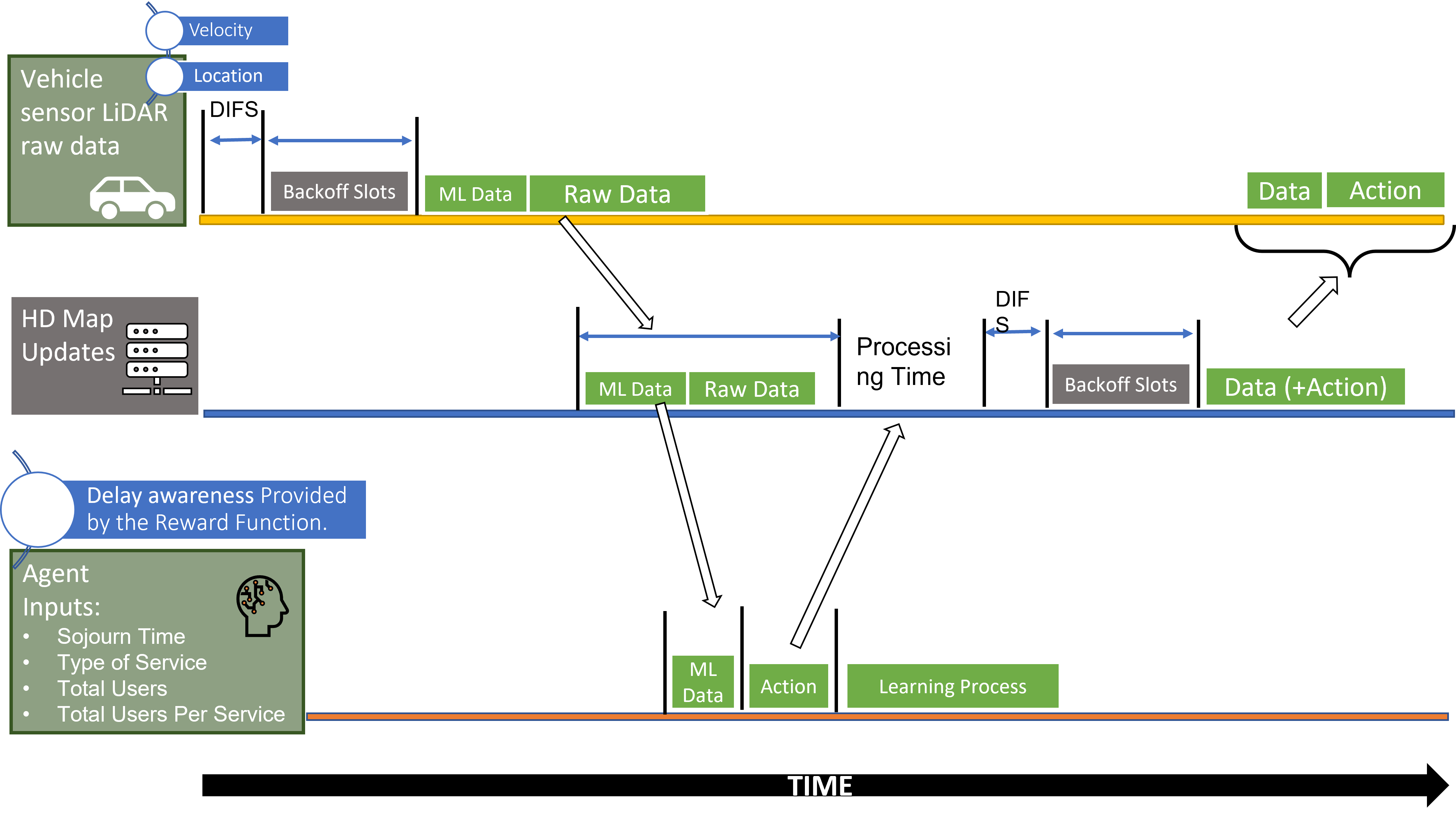}
  \caption{Data Flow, in time domain, from Vehicle to Agent to Vehicle.}
  \label{fig:TimeFlow}
  \end{center}
\end{figure}

The waiting time strategy is summarized in Algorithm 3 and 4. Refer to Fig. \ref{fig:TimeFlow} for data flow between vehicle and agent.

\begin{algorithm}
\caption{Q-Learning}
\label{algo:q_learning}
\begin{algorithmic}[1]
\STATE Initialize the agent with epsilon $\epsilon
$ , number of actions $k$.
\
\FOR {number\_of\_episodes}
\WHILE{STATUS}
  \STATE get observation of the environment
  \STATE RSU extracts Velocity, Location
  \STATE RSU calculates Sojourn Time
  \STATE RSU provides $\mathcal{S} = \{S_j , T_v , C, T_cv\}$
  \STATE agent select $a = $ choose\_action(S) Comment: Algorithm (\ref{algo:choose_action})
  \STATE agent sent action to $v$ through RSU
  \STATE RSU maps the actions with eq. (\ref{eq:map_actions})
  \STATE RSU calculates the $r_t$ according to the Utility function (\ref{eq:utility_function_penalty}), penalties and bonuses (\ref{eq:penalties_rate_VO_VI_HD}, \ref{eq:penalties_BE1}, \ref{eq:bonus_Latency_VO_VI_HD_BE}), and thresholds tables (\ref{tab:data_rate_thresholds}), (\ref{tab:delay_thresholds}) 
  \
  \STATE agents received next\_observation, $r_t$, STATUS
  \STATE update $Q(s,a)$ by using eq. (\ref{eq_q_learning_2})
\ENDWHILE
\ENDFOR
\STATE \textbf{Finish the subroutine}
\end{algorithmic}
\end{algorithm}

\begin{algorithm}[h]
\caption{choose\_action(S)}
\label{algo:choose_action}
\begin{algorithmic}[1]
  \STATE c = extract\_category(S)  
      \IF {$p(\epsilon) < \epsilon$}
        \STATE $a^* = random(\mathcal{A})$ 
      \ELSE
        \STATE $a^* = \arg\max_a Q(s, a)$
      \ENDIF
  \STATE $w = a \cdot w_{max}(c)/|\mathcal{A}|$
\RETURN $w$
\end{algorithmic}
\end{algorithm}

%% file: table_simulation.tex
\section{Simulation} \label{simulation}
In order to evaluate the efficacy of our RL solution, we utilized OMNet++ version 6 \cite{omnetppOMNeTDiscrete} and Inet Framework \cite{omnetppINETFramework}. These tools were chosen due to their ability to provide the IEEE802.11p protocol stack for wireless communication. We also incorporated Veins \cite{veins} to facilitate communication between OMNet++ and SUMO \cite{SUMO} as our mobility model. Finally, we integrated and adapted the tool developed by M. Schettler et al \cite{veinsgym} to implement our machine learning algorithm in Python, in conjunction with the inet framework. The simulation scenario consists in a tile, one RSU in the middle of it, and vehicles moving within the tile. Please refer to Table \ref{table:simulation_parameters} for a detailed breakdown of our simulation parameters. 

\subsection{Traffic Flow Generation}
The traffic flow was generated from a real dataset provided by Traffic and Accident Data Unit / North East Regional Road Safety Resource \cite{dataset}. The insertion of vehicles to the simulation, entry interval time, was calculated by the total average of vehicles in a rush hour (12 p.m.) divided by one hour. Thus, during one rush hour with $2,376$ vehicles, we have an entry interval time of $0.66$s. Additionally, the AVs enter and leave the simulation environment following a random trajectory. 

\subsection{Data Assignment} \label{section:data_assignment}
Whenever a vehicle enters the simulation environment, the application will select a category randomly from the set $C =\{VO,VI,BE,HD\}$. After that, the AV transmits that particular type of data to the RSU. Each of the categories generates data with a rate stated in Table \ref{table:simulation_parameters}. For voice G.711 the data rate is between $80\text{kbps}-100\text{kbps}$ \cite{dataRate_cisco_voice_szigeti2005end}, for a video of 720p is $5\text{Mbps}$ \cite{dataRateVideo_googleYouTube}, for HD Map is $4\text{Mbps}$ \cite{5gaa_delay}, and for best effort that consists of all other applications, it is saturated with $28\text{Mbps}$. With random generation of trajectories and category assignment, we can secure that the algorithm is not biased.

\begin{table}[ht]
\caption{Simulation Parameters}
\begin{center}
\begin{tabular}{|c|c|}
\hline
\textbf{\textit{Parameter}} & \textbf{\textit{Value}} \\
\hline
Tile Dimension & 300x100 meters  \\
\hline
Episodes & 50  \\
\hline
Episode duration & 250s  \\
\hline
$\epsilon-greedy$ & 0.2  \\
\hline
Discount Factor $\gamma $& 0.99  \\
\hline
Learning Rate $\epsilon $& 0.1  \\
\hline
$\alpha_1 $, $\alpha_2 $ & 0.3 and 0.7 respectively  \\
\hline
Simulation time & 250s  \\
\hline
Vehicle Density & varies according traffic flow  \\
\hline
Coverage Area & 200m  \\
\hline
Vehicle max speed & 17m/s  \\
\hline
Vehicle Acceleration  & {2.6 m}/{$s^2$} Average value of petrol cars\cite{acceleration_deceleration_petrol_car}. \\
\hline
Vehicle Deceleration  & {4.5 m}/{$s^2$} Average value of petrol cars\cite{acceleration_deceleration_petrol_car}.  \\
\hline
Tx Power & 200 mW  \\
\hline
Frequency & 5.9 GHz  \\
\hline
Bandwidth & 10MHz  \\
\hline
Best-effort data rate & 28Mbps  \\
\hline
HD Map data rate & 4Mbps \\
\hline
Video data rate & 5Mbps  \\
\hline
Voice data rate & 100kbps  \\
\hline
TXOP limit & Disabled as per standard  \\
\hline
\end{tabular}
\label{table:simulation_parameters}
\end{center}
\end{table}

\subsection{Actions, Thresholds and Rewards}

\paragraph{Actions}
After conducting a thorough analysis of the data and extensive experimentation, we have determined that eight actions are optimal. We advocate for adopting the waiting time values from Table \ref{tab:maxi_actions} to ensure stability and priority, aligning with the standard EDCA. These chosen values are set below the maximum latency in the Cumulative Distribution Function (CDF) in Fig. \ref{fig:cdf_latency}, where the standard EDCA is represented by the line ``IEEE802.11p without AC HD Map (standard EDCA)''. This strategy has led to significant improvement, detailed in the results and discussion section \ref{results}. 

\begin{table}[h]
\centering
\caption{Maximum Waiting Time per Category}
\begin{tabular}{c|c}
   \hline
   \multicolumn{2}{l}{Maximum Waiting Time} \\
   \hline
   VO  & 0.92s \\
   VI  & 2s \\
   HD Map  & 2s \\
   BE  & 8s \\
\end{tabular}
\label{tab:maxi_actions}
\end{table}

By selecting the same maximum waiting time for HD Map equal to VI, we aim to share the communication channel equally without affecting the priority VI has over HD Map. 
\\
\paragraph{Thresholds}
To achieve optimal results, it's crucial to select the appropriate QoS configuration adjusting penalties and bonuses as necessary to prioritize. As such, we've established thresholds in line with standard requirements: for voice, we have set the data rate threshold at $100$kbps \cite{dataRate_cisco_voice_szigeti2005end}; for video, we have decided that the system should support a minimum resolution of $360$p in worst-case scenarios \cite{dataRateVideo_googleYouTube}, and set the threshold at $1.25$Mbps accordingly. For HD Map, we've opted for the same threshold as video to maintain balance among services.
Lastly, BE AC has the lowest priority and incurs a penalty if the data rate exceeds $1$Mbps (as per penalty eq. (\ref{eq:penalties_BE1})). By setting these thresholds, we ensure that each service maintains its desired priority, and you can find specific details in Tables \ref{tab:data_rate_thresholds} and \ref{tab:delay_thresholds}.
Regarding the delay threshold, we have followed the same analogy.
\\
\paragraph{Rewards}
Based on the significance of enhancing latency in our system, without compromising throughput. We conducted exhaustive simulations to determine the optimal allocation of weights to the utility function's parameters. Our analysis concluded that the most effective approach involves assigning weights of 0.3 and 0.7 for $\alpha_1$, $\alpha_2$ respectively.

\begin{table}[h]
\begin{minipage}{.25\textwidth}
\centering
\caption{Data Rate Thresholds}
\begin{tabular}{c|c}
   \hline
   \multicolumn{2}{l}{Data Rate Thresholds} \\
   \hline
   VO & 100kpbs \cite{dataRate_cisco_voice_szigeti2005end}\\
   VI & 1.25Mbps \cite{dataRateVideo_googleYouTube}\\
   HD Map & 1.25Mbps \\
   Best-Effort & 1.0Mbps \\
\end{tabular}
\label{tab:data_rate_thresholds}
\end{minipage}%
\begin{minipage}{.25\textwidth}
\centering
\caption{Delay Thresholds}
\begin{tabular}{c|c}
   \hline
   \multicolumn{2}{l}{Delay Thresholds} \\
   \hline
   VO  & 150 ms \cite{latency_cisco} \\
   VI  & 100 ms \cite{dataRate_cisco_voice_szigeti2005end}\\
   HD Map  & 100 ms \cite{5gaa_delay}\\
   Best-Effort  & 1000 ms \\
\end{tabular}
\label{tab:delay_thresholds}
\end{minipage}
\end{table}

Accordingly, we conducted extensive simulation to determine the best values for penalty and bonus ensuring that the voice category is given top priority and the BE category the lowest. Consequently, the values of minus two for VO, VI, HD; and minus ten for BE were selected. For bonuses, all services received plus two, see equations (\ref{eq:penalties_rate_VO_VI_HD}), (\ref{eq:penalties_BE1}), and (\ref{eq:bonus_Latency_VO_VI_HD_BE}).

\subsection{Key Performance Indicator (KPI)}
For the analysis of the simulation, three main KPIs are considered: latency, throughput, and fairness. Latency is calculated by the difference between generated and received packet time. The throughput is calculated by the total of packets received within a specific time frame. For fairness, the formula to express fairness is Jain's fairness index \cite{fairness_formula_jain}, see eq. (\ref{eq:fairness}). 

\begin{equation}
    J(x_1,x_2,...,x_n) = \frac{(\sum_{i=1}^n x_i)^2}{n \cdot \sum_{i=1}^n x_{i}^2}
    \label{eq:fairness}
\end{equation}

\subsection{Exploration-exploitation, discount factor and learning rate}
For RL, it is imperative to carefully select an strategy to balance exploration and exploitation. In this study, specifically, we aim for the agent to exert minimal influence on the network. Following this approach, the system will be optimal according to the standard throughout the learning process. Therefore, utilizing an $\epsilon$  of $0.2$ ensures that the system commences with minimal intervention from the agent.

Following several simulations for fine-tuning, we decided to set the hyperparameters with a discount factor of 0.99 and a learning rate of 0.1. The rationale behind these choices lies in our emphasis on future rewards. Having a learning rate of 0.1 has been found to be effective to avoid having many unvisited states driven by the high mobility of vehicles.

%% file: results.tex
\section{Result and Discussion} \label{results}

Results of the proposed RL solution are compared with the current standard IEEE802.11p without QoS, with QoS, with the new AC HD Map, and the single agent without penalties.

\subsection{Comparison Against Current Standard } 

The proposed Sojourn Time Q-Learning algorithm showed an overall improvement in latency.

\subsubsection{Latency}
\begin{figure}[h]
	\centering
	\includegraphics[width=\linewidth]{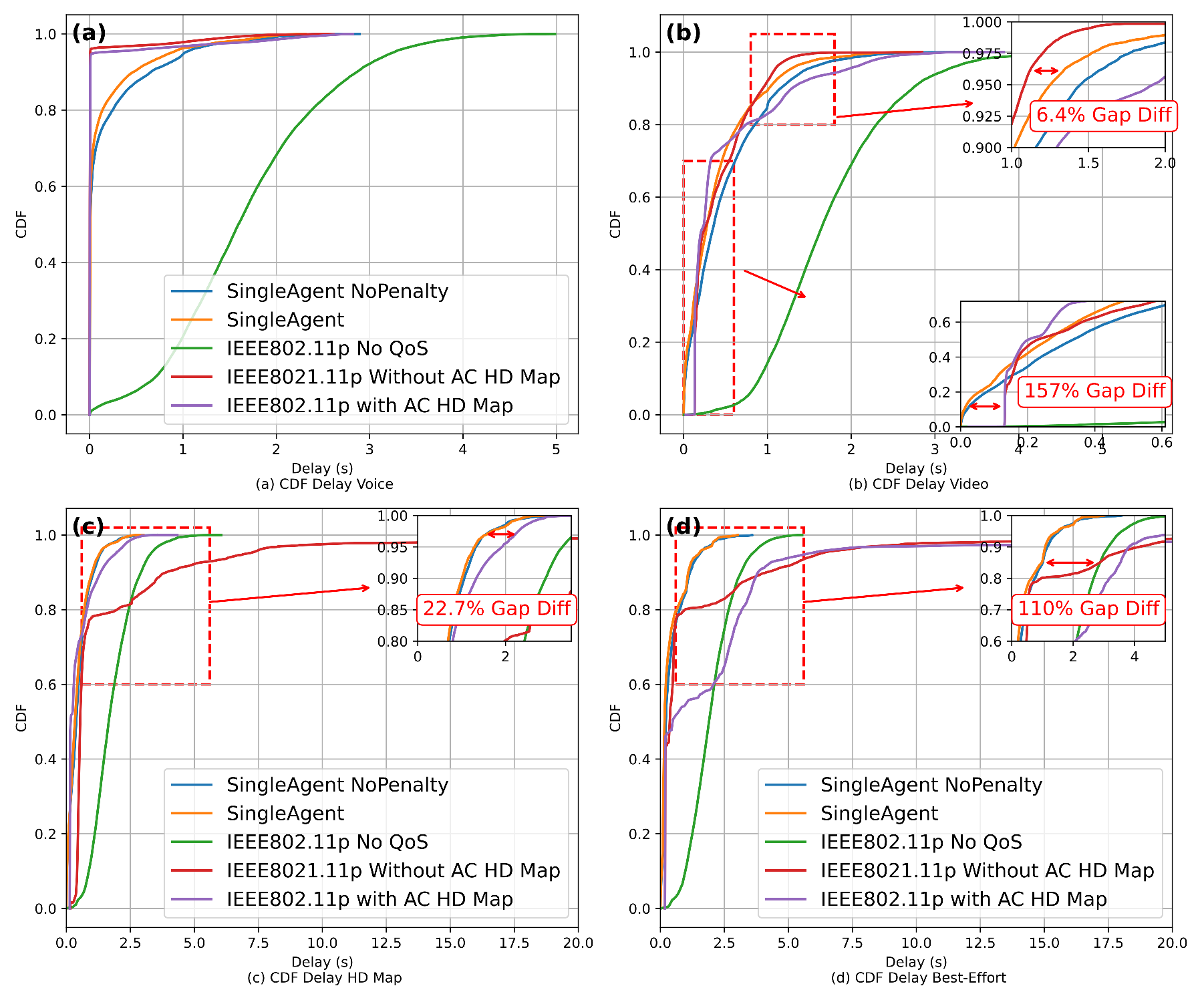}
	\caption{Latency Cumulative Distribution Function (CDF) (a) voice, (b) video, (c) HD map, (d) best-effort (Standard IEEE802.11p).}
	\label{fig:cdf_latency}
\end{figure}

\begin{figure}[h]
	\centering
	\includegraphics[width=\linewidth]{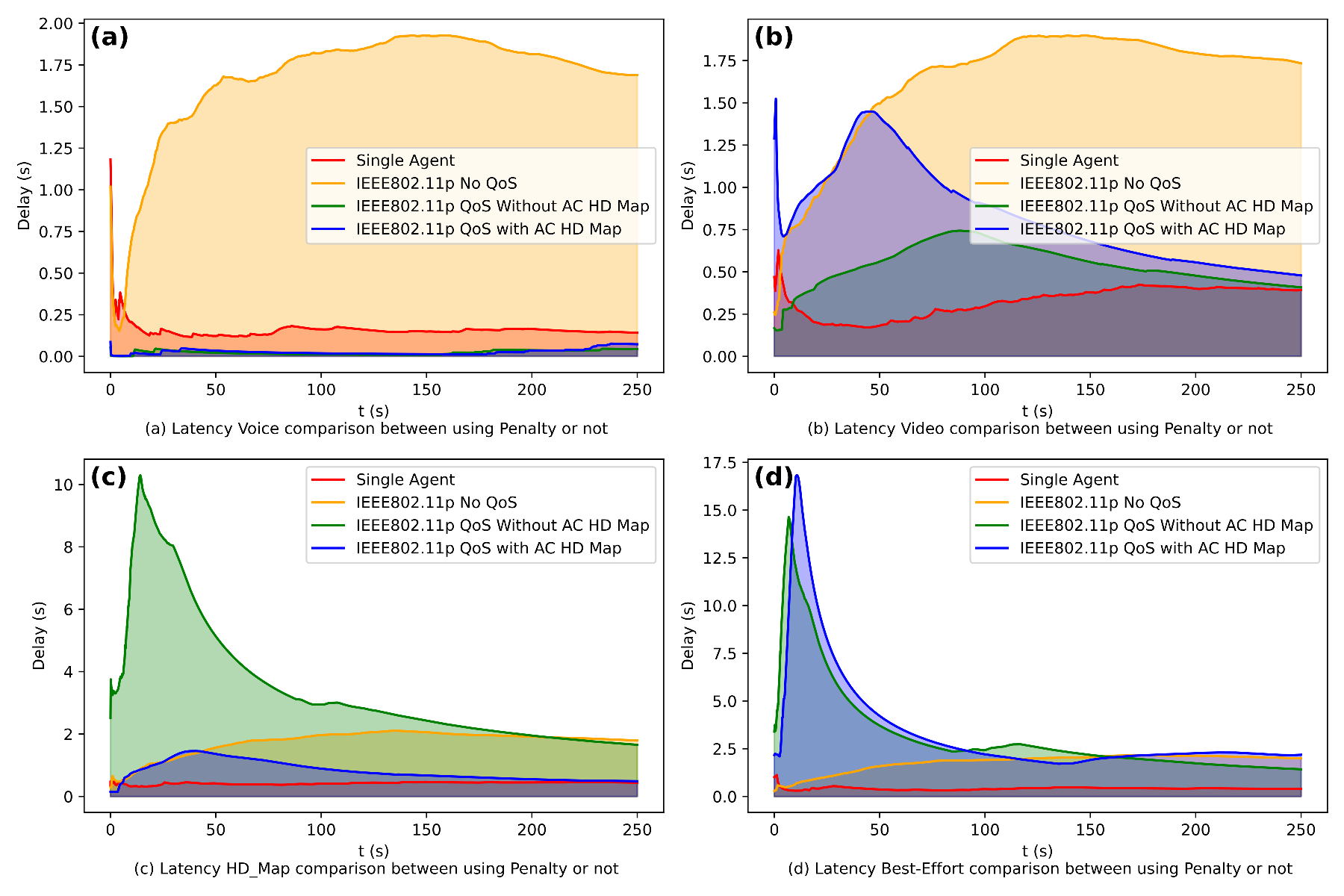}
	\caption{Latency comparison in time domain (a) voice, (b) video, (c) HD map, (d) best-effort (Standard IEEE802.11p).}
	\label{fig:time_latency}
\end{figure}

In Fig. \ref{fig:cdf_latency}, despite not enabling EDCA, the performance of a single agent exceeds IEEE802.11p when no penalties are applied. There is a remarkable reduction in latency with values of $92\%$, $77\%$, $75\%$, and $80\%$ for voice, video, HD map, and best-effort respectively. While using EDCA and EDCA with the new HD map AC, the delay for voice is lower compared to our solution, see Fig. \ref{fig:cdf_latency}(a). However, the agent attempted to maintain the latency under the threshold stipulated of $150$ms as seen in Fig. \ref{fig:time_latency}(a). 

For the other three services, the agent was able to generate a lower average latency against IEEE802.11p EDCA with reduction of $2.5\%$ for video, $73\%$ for HD map, and $73\%$ for best-effort. Compare with the new HD Map AC, there is a decrease of $18.5\%$ for video, $10\%$ for HD map, and $82\%$ for best-effort. In Fig. \ref{fig:cdf_latency}(b), for video, there are two areas to evaluate. For the part where latency is lower, it is clear that the single agent can reach lower values with a gain of $157\%$ compared with the IEEE802.11p EDCA and new AC HD Map. As the Cumulative Distribution Function (CDF) approaches 1.0, the standard IEEE802.11p EDCA was able to reach lower delay with a marginal deviation of 6.4\%. However, as seen in Fig. \ref{fig:time_latency}(b), the agent maintained a lower average delay. 

For HD Map, it is seen in Fig. \ref{fig:cdf_latency}(c), and \ref{fig:time_latency}(c) that the agent kept the latency lower than the standard and new AC category. The improvement was $75\%$, $73\%$, and $10\%$ compared with IEEE802.11p No QoS, IEEE802.11 with QoS, and IEEE802.11p with new AC for HD map, respectively. For the CDF range between 0.8 and 1.0, the agent showed an improvement between $14\%$ to $22.7\%$. Concerning the best-effort category, as shown in Fig. \ref{fig:cdf_latency}(d), our solution outperforms the other approaches across the range from $0.6$ to $1.0$, showing an improvement of 110\%.

\begin{figure}[h]
	\centering
	\includegraphics[width=\linewidth]{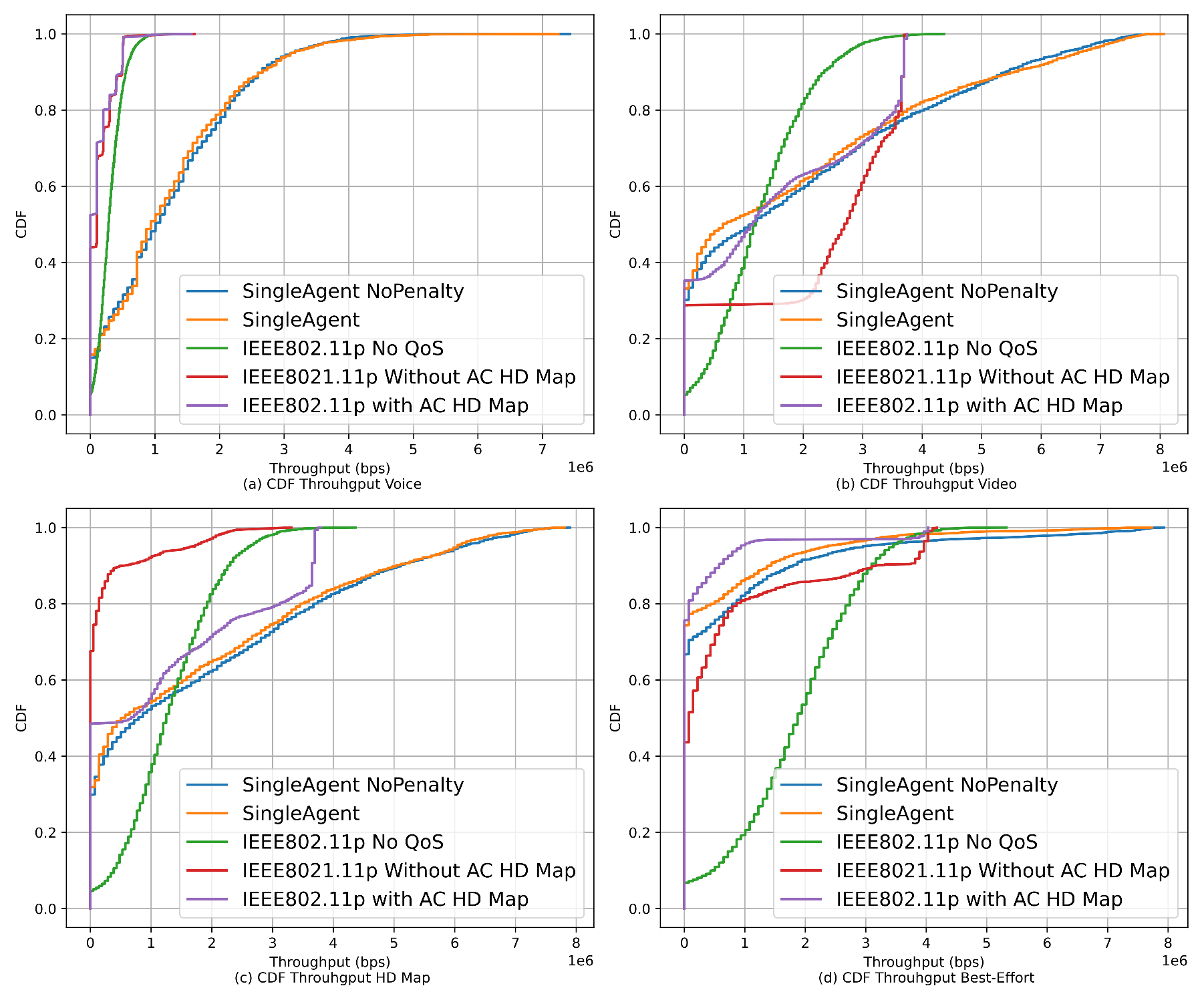}
	\caption{Throughput CDF (a) voice, (b) video, (c) HD map, (d) best-effort (Standard IEEE802.11p).}
	\label{fig:cdf_throughput}
\end{figure}

\subsubsection{Throughput}
\begin{figure}[h]
	\centering
	\includegraphics[width=\linewidth]{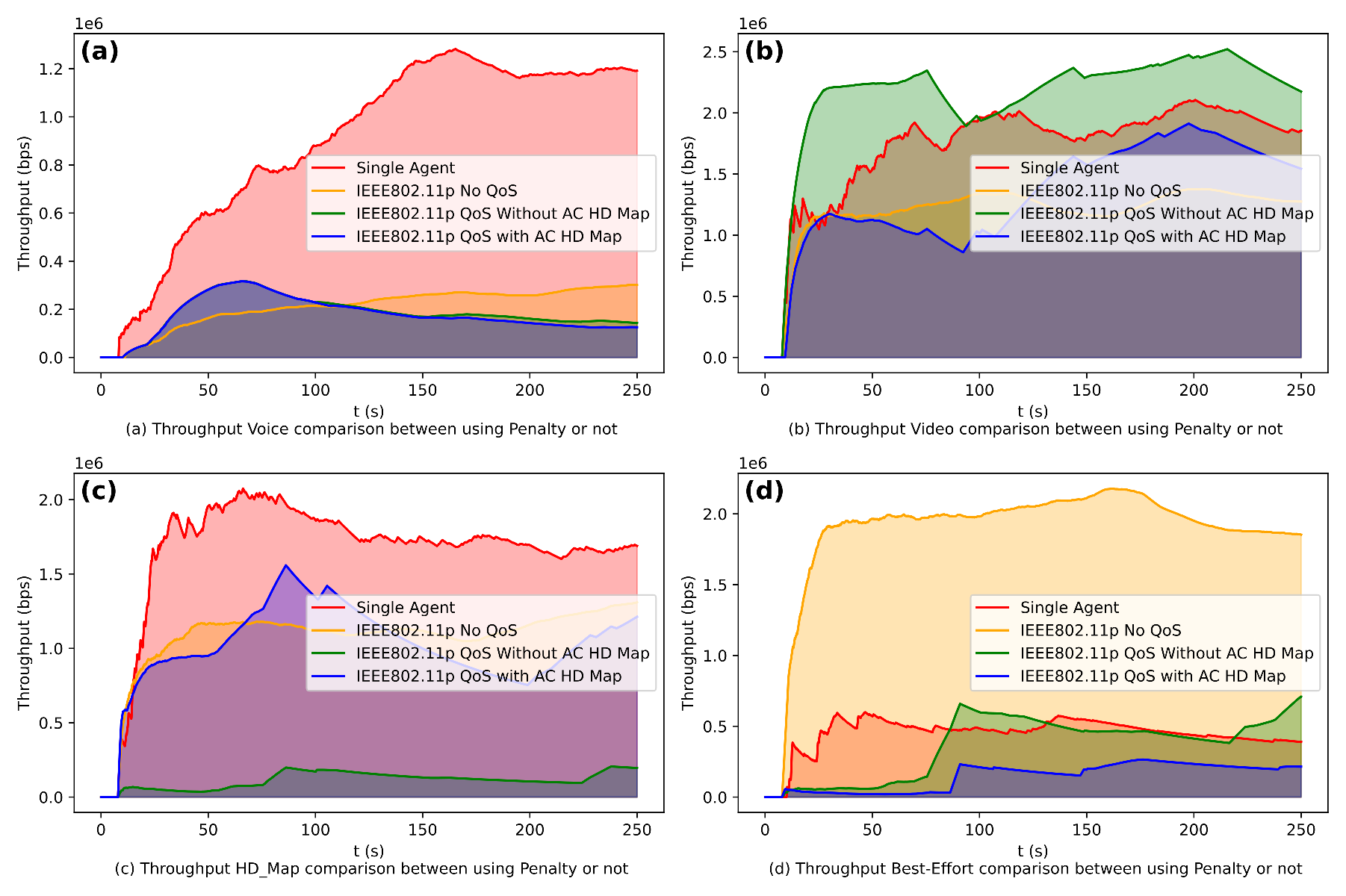}
	\caption{Throughput comparison in time domain (a) voice, (b) video, (c) HD map, (d) best-effort (Standard IEEE802.11p).}
	\label{fig:time_throughput}
\end{figure}

For the throughput, in the CDF Fig. \ref{fig:cdf_throughput}(a), \ref{fig:cdf_throughput}(b), and \ref{fig:cdf_throughput}(c) is observed the line corresponding to the single agent is placed to the rightmost for voice, video, and HD map indicating a higher data rates, with maximum value of approximately $7$Mbps. For the BE Fig.\ref{fig:cdf_throughput}(d) that is located to the left which is the desired outcome as it is the lower priority.
In Fig. \ref{fig:time_throughput}(b), and \ref{fig:time_throughput}(c), it is observed how the single agent can achieve a higher stability and maintain the throughput above the threshold $1.25$Mbps for video, and HD map data. Voice has an increase of $300\%$ compared with the standard. For video compared with the new AC HD Map, the increase is $15\%$, and compared with the normal EDCA there is a decrease of $33\%$. As per HD map data, there is an increase of $33\%$ compared with the standard and new AC.

\subsubsection{Fairness}
\begin{figure}
		\centering
		\includegraphics[width=3in]{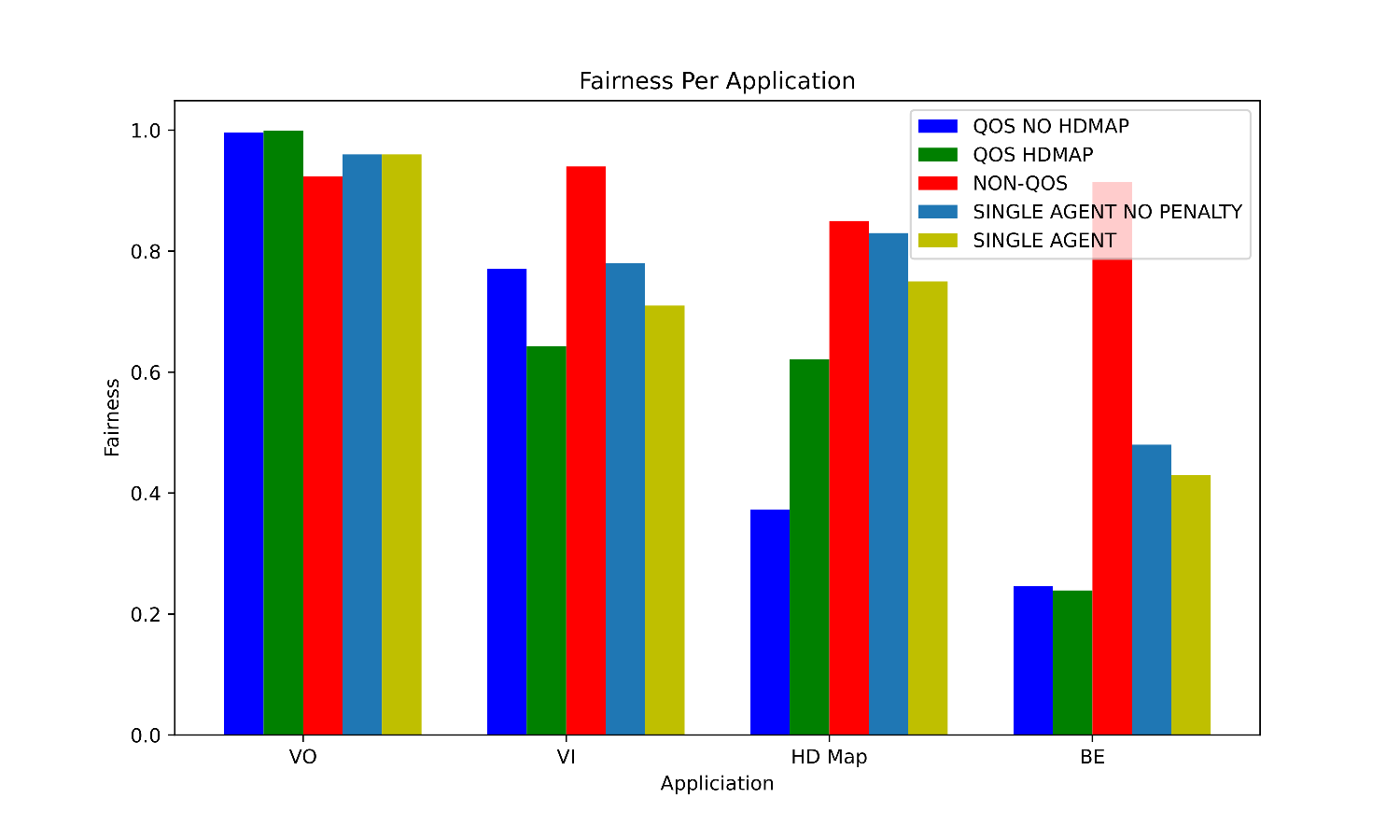}
		\caption{Fairness comparison per service between IEEE802.11p, EDCA, New AC HD Map, single agent with and without penalties.}
		\label{fig:fairness}
	\end{figure}

Fig. \ref{fig:fairness} reveals that the voice for all solutions is about $0.95$ which is a high index. For other services, the single agent with respect to voice and video is able to perform similarly to the EDCA and the new AC category HD Map. The main difference resides in the HD Map and best effort. By comparing against the new AC HD Map, our solution increased fairness by $36.9\%$, $32\%$, and $113.4\%$ for video, HD map, and best-effort respectively.

	\subsection{Validation of the model} 
	We conducted a validation tests on the model using the IEEE802.11a standard. 
	
	\subsubsection{IEE802.11ac}
\begin{figure}[h]
		\centering
		\includegraphics[width=\linewidth]{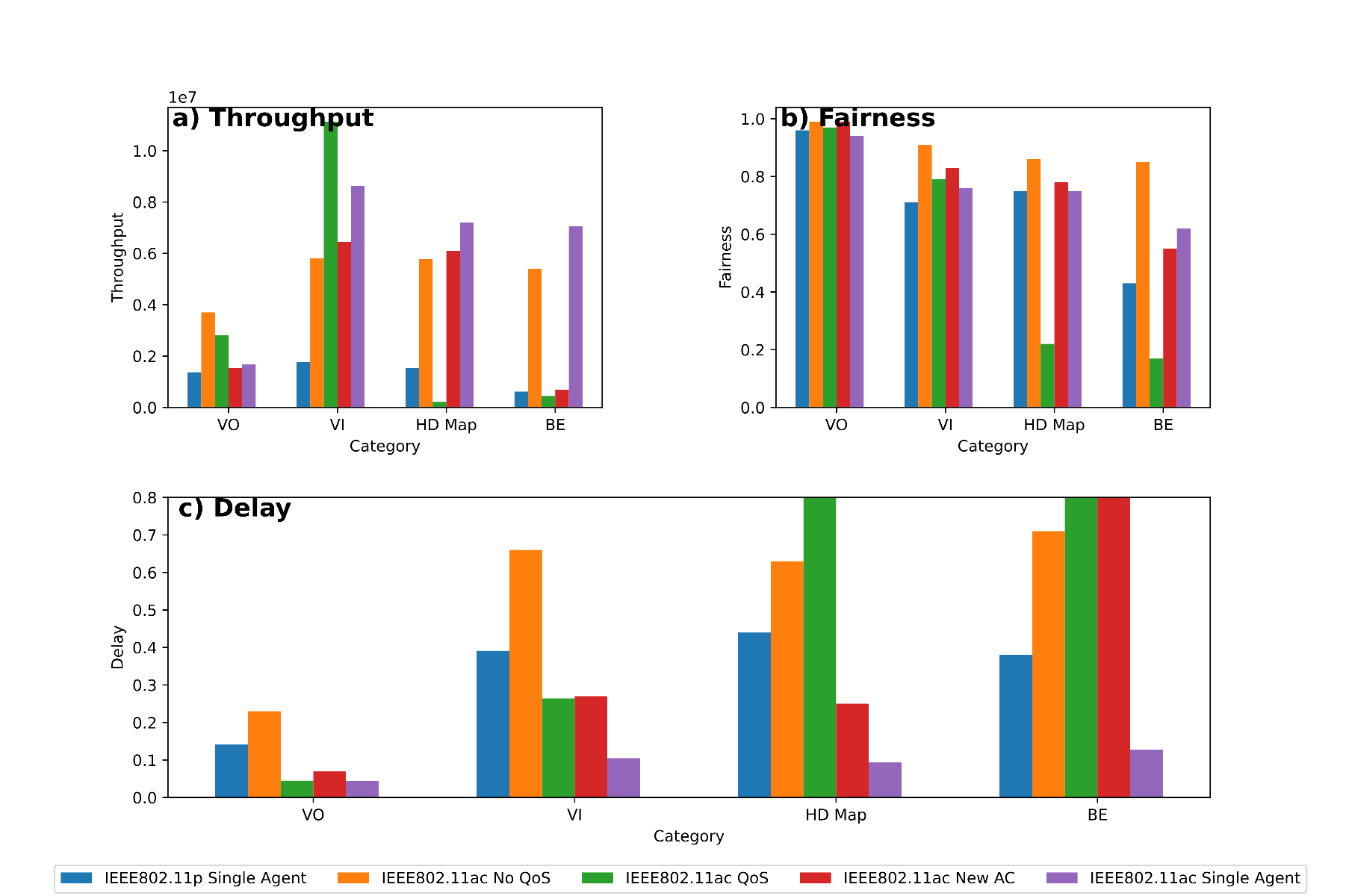}
		\caption{Comparison between IEEE802.11p and IEEE802.11ac standard using same learned model. The subplot (c) the y-axis limit was set to 0.8 to appreciate the delay difference due to the high delay value of HD map and BE (6s, and 3s, respectively).}
		\label{fig:ieee802_11_comparison}
	\end{figure}

	IEEE802.11ac was considered for further evaluation. As expected, it provided a higher throughput and lower latency. For our model, as it is seen in Fig. \ref{fig:ieee802_11_comparison}(c) the single agent was able to provide almost the same latency as EDCA for voice with only $0.4\%$ difference in the average delay. For video, the single agent improved the delay by $60\%$, and $105\%$ for HD Map.
	
	Whilst comparing with the new AC approach, the single agent was able to ou.tperform by $36\%$, $61\%$, $64\%$, and $93\%$ for voice, video, HD Map, and best-effort respectively. Regarding Throughput (Fig. \ref{fig:ieee802_11_comparison}(a)), the single agent was able to provide higher throughput with an increase of $32\%$, $14\%$, and $600\%$ for video, HD map, and BE, respectively. When compared to IEEE802.11ac QoS, the single agent presented an average throughput lower than the EDCA with respect to voice and video. This can be attributed to the prioritization of the aforementioned services. However, the conventional EDCA approach restricts the throughput for HD map which we are looking to optimize. It shows that the proposed Q-learning has an adequate fairness index providing lower delay and keeping throughput stable for HD Map and Video. In addition, it allows the BE (lower priority) category to utilize the spectrum with a higher quality than the current EDCA.

%% file: conclusion.tex
\section{Conclusion} \label{conclusion}
In summary, this paper proposed an RL Q-learning that incorporates sojourn time as part of its state to improve QoS. It provides coverage awareness to the agent. The key feature of this solution is its ability to enhance QoS by mimicking the backoff process of the standard IEEE802.11 EDCA and improving it without requiring any changes. The results demonstrated a significant improvement in latency for HD map data, with percentages of $75\%$, $73\%$, and $10\%$ compared with IEEE802.11p without Quality of Service (QoS), IEEE802.11 with QoS, and IEEE802.11p with new access category (AC) for HD map, respectively. Additionally, results showed an improvement when the solution was evaluated with the standard IEEE802.11ac, this demonstrated that our solution overcomes the problem of compatibility. Moreover, while using the standard IEEE802.11ac, it is observed a better performance in terms of throughput and latency, highlighting the insufficiency of the standard IEEE802.11p to support low-latency and heavy-data traffic applications such as HD Map. Recognizing these limitations, future work may necessitate exploring a newer wireless technology and implementing dynamic packet size strategies to optimize Quality of Service (QoS). In addition, it is important to acknowledge that this solution is based on a single-agent solution. As the complexity of the environment increases, future endeavors may call for the adoption of multi-agent approaches.